# Integrating Taxonomies into Theory-Based Digital Health Interventions for Behavior Change: A Holistic Framework

Yunlong Wang[1], Ahmed Fadhil[2], Jan-Philipp Lange[3], Harald Reiterer[1]

[1]HCI Group, University of Konstanz, Konstanz, Germany, {firstname.surname}@uni-konstanz.de

[2]ICT4G Group, Fondazione Bruno Kessler, Trento, Italy, fadhil@fbk.eu

[3]Social and Health Sciences Group, University of Konstanz, Konstanz, Germany, jan-philipp.lange@uni-konstanz.de

## Abstract

Digital health interventions have been emerging in the last decade. Due to their interdisciplinary nature, digital health interventions are guided and influenced by theories (e.g., behavioral theories, behavior change technologies, persuasive technology) from different research communities. However, digital health interventions are always coded using various taxonomies and reported in insufficient perspectives. The inconsistency and incomprehensiveness will bring difficulty for conducting systematic reviews and sharing contributions among communities. Based on existing related work, therefore, we propose a holistic framework that embeds behavioral theories, behavior change technique (BCT) taxonomy, and persuasive system design (PSD) principles. Including four development steps, two toolboxes, and one workflow, our framework aims to guide digital health intervention developers to design, evaluate, and report their work in a formative and comprehensive way.

## Introduction

According to the County Health Rankings [46], variation in health can be accounted for by health behaviors (30%), clinical care (20%), social and economic factors (40%), and physical environment (10%). Increasing evidence shows that lifestyle-related behaviors, such as diet, exercise, sleeping, emotion, and smoking play an essential role in people's health. Chronic diseases caused by unhealthy behaviors and habits are among the leading causes of mortality [24]. Some of the chronic diseases, e.g., type 2 diabetes, could be life-long and bring a heavy burden to the patients and their family. The only way to prevent many chronic diseases is to change unhealthy lifestyles, e.g., diet and physical activity.

With the potential for low cost and high scalability for chronic disease prevention, in the past decade, digital health interventions (DHIs) have been widely discussed by government stakeholders, clinicians, and researchers [60]. Designing and deploying DHIs are challenging due to the complexity of human behavior, which could be affected by individuals' motivation, emotion, ability, social environment, and physical environment. Therefore, DHI design could accordingly require theories and practice from several disciplines, including phycology, public health, behavioral science, human-computer interaction, and so on. The interdisciplinary nature of DHIs calls for a comprehensive framework to guide the development, evaluation, and report.

As DHIs are expected to change human behavior, behavioral theories can serve as the development foundation. It has shown that theory-based behavior change interventions are more effective than others [15,23]. Nevertheless, behavioral theories could also be ignored [58] or misused [25]. Although behavioral theories allow to explain and predict behavior, they lack the guidance of translating into operational techniques.

The Behavior Change Technique (BCT) taxonomy [1] and persuasive technology design (PSD) principles [42] are two widely used taxonomy in DHIs research [21,27,35,43]. These taxonomies not only inform DHI design but also enable precise reporting, which will be favored by systematic reviewers. Although derived from different philosophies, BCTs and PSDs have some common elements. However, they are used separately in many DHI studies. To benefit from both, we combine the BCT taxonomy and PSD principles into a more comprehensive taxonomy in the light of the Behavioral Intervention Technology (BIT) model [36].

In this paper, we aim to put the puzzles together and build a holistic framework to aid DHIs researchers to design, evaluate, and report their studies. In short, our contributions include:

(1) We provide a holistic framework that allows DHI developers to design, evaluate, and report their work in a formative and comprehensive way.
(2) We classified PSD principles into two parts: strategies and characteristics. We then combine the BCT taxonomy and PSD principles (the characteristics) into our DHI taxonomy.
(3) By elaborating the BIT model, we propose a comprehensive way to report DHI description: strategies, characteristics, and a workflow.

# Related Work

As this paper is for DHI developers from different communities, it is necessary to clarify the terms and our scope before we present the related work. *Digital health* or *eHealth* is the umbrella concept referring to the use of information and communication technologies (ICT) for health [61]. According to the world health organization (WHO), *digital health interventions* (DHIs) covers systematic functionalities to support clients, healthcare providers, health system or resource managers, and data services [60]. In this paper, however, we limit our scope to the DHIs aiming to change users' lifestyle behavior (e.g., food intake, physical activity, and smoking) using digital technology to prevent or manage health problems.

## CeHRes Roadmap

Back in 2011, a holistic framework (i.e., CeHRes Roadmap) was proposed to improve the uptake and impact of eHealth technology. The CeHRes roadmap was built upon 16 existing frameworks via a systematic review and emphasized the importance of holism [20]. Human characteristics, socioeconomic and cultural environments, and technology are closely connected to affect human behavior. Therefore, developers should always keep these holistic factors in mind to build eHealth technologies. Within this framework, CeHRes roadmap was illustrated as a practical guideline to help plan, coordinate, and execute the participatory development process of eHealth technologies. CeHRes roadmap consists of six steps - contextual inquiry, value specification, design, operationalization, and summative evaluation - which integrate persuasive technology design, human-centered design, and business modeling. Although CeHRes roadmap integrates behavioral theories as for the foundation, it does not explicitly show how to apply them in the intervention design. Besides, CeHRes roadmap does not adopt any persuasive technology taxonomy.

## Behavioral Intervention Technology Model

In 2014, Mohr and colleagues proposed the behavioral intervention technology (BIT) model aiming to support the translation of treatment and intervention-aims into an implementable treatment model [36]. The BIT model includes a theoretical phase followed by an instantiation phase. The theoretical phase consists of the intervention aims and behavior change strategies, whereas the instantiation level consists of intervention elements, characteristics, and workflow. Thus the BIT model can serve as a supplement to the CeHRes roadmap. However, the BIT model only provided some examples in each component. E.g., behavior change strategies include education, goal setting, monitoring, feedback, and motivation enhancement. As the author mentioned, the BIT model is a simplification and should be modified and elaborated to fit users' need [36]. In this paper, we will adjust and elaborate the BIT model to fit into our holistic framework.

### IDEAS

In 2016, Mummah et al. proposed IDEAS (Integrate, Design, Assess, and Share) as a framework and toolkit of strategies for the development of DHIs [39]. IDEAS was built on three essential components: behavioral theory, design thinking, and evaluation and dissemination. The IDEAS framework emphasizes the importance of behavioral theories and introduces the taxonomy of behavior change techniques (BCTs). However, the BCT taxonomy is regarded as an alternative to using behavioral theories to identify target constructs in interventions. In our holistic framework, we suggest using both of them as two necessary steps because they correspond to the intervention aims and strategies respectively.

## Behavioral Theories

All the three reviewed work above mention behavioral theories, but only IDEAS explicitly integrate behavioral theories into the step development process. Behavioral theories refer to the social-psychological theories of behavior change, which explain and predict human behavior. As depicted by Sutton [50], each of the behavioral theories specifies a small number of cognitive and affective factors as the proximal determinants of behavior (see Figure 1). These factors are called *constructs* in behavioral science [25]. We will use this term to refer to the fundamental components of behavioral theories in the rest of the paper.

Glanz et al. [22] illustrated the most frequently used behavioral theories published before 2010: the Social Cognitive Theory (SCT) [6], the Transtheoretical Model of Change (TTM) [44], the Health Belief Model (HBM) [48], and the Theory of Planned Behavior (TPB) [2]. Davis et al. [45] also identified 82 behavioral theories, among which the most frequently used theories are TTM, TPB, SCT, the Information-Motivation-Behavioral-Skills Model, HBM, the Self-determination Theory [49], the Health Action Process Approach (HAPA) [53], and the Social Learning Theory [5]. Based on different assumptions of human behavior, these behavioral theories can be grouped into continuum theories and stage theories [11].

Continuum theories assume people's behavior is caused by a set of variables, e.g., intention and skills. Except for TTM, all other mentioned theories fall into this group. Based on the behavioral model integrating several existing ones [11], we present a hypothesized continuum model as shown in Figure 1. Planning (shown in red in Figure 1) is specified as a mediator of the intention-behavior relationship in HAPA [53,55]. Habit (shown in green in Figure 1) has been found being able to moderate the effects of planning [32].

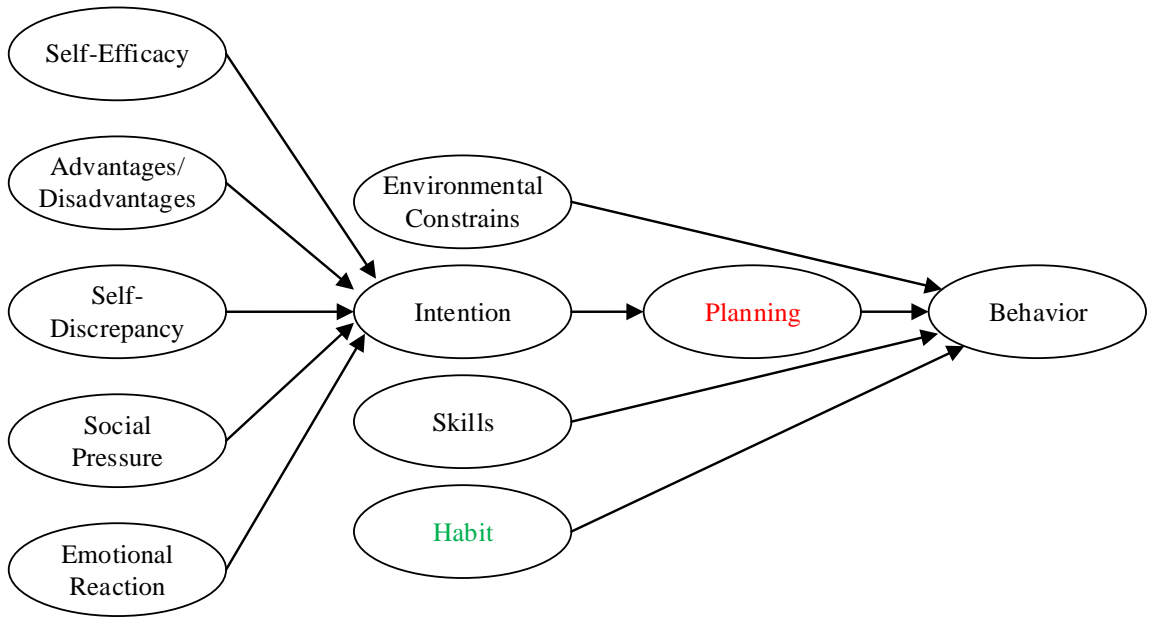

Figure 1: A hypothesized continuum model. The constructs in black are borrowed from the integrated behavioral model in [11]. The construct "planning" is from the Health Action Process Approach (HAPA) [54], and the construct "habit" is added inspired by the work [32].

Stage theories assume people change their behavior in a process including several stages. The factors pushing people from one stage to the next are believed to be different. Therefore the strategies at each state should be adapted accordingly. E.g., Figure 2 shows the stages and strategies of TTM.

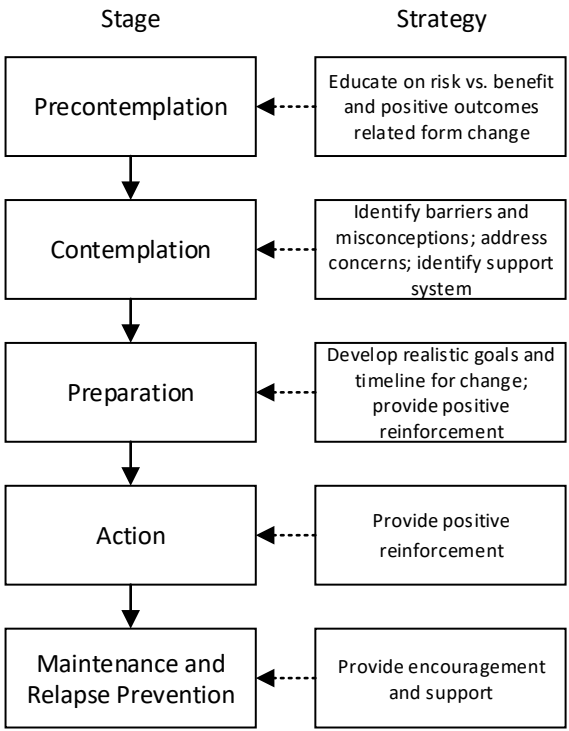

Figure 2: The Transtheoretical Model of Change (TTM), adapted from [4]. This model divides the behavior change process into five stages, namely *precontemplation, contemplation, preparation, action, and maintenance and relapse prevention.* Depending on the stage of change, different strategies could be applied accordingly to make the intervention effective.

Behavioral theories provide a toolbox to understand human behavior and explain the rationale behind interventions. However, their shortcomings should be noticed before they are used. Hekler and colleagues [25] have pointed out three shortcomings of behavioral theories: (i) most behavioral theories explain only a small portion of variance in the outcomes they are trying to account for; (ii) many behavioral theories, in their current form, are not falsifiable; and (iii) there is a fragmentation and an over-abundance of different theories. Therefore, DHI developers should base on without being limited to behavioral theories. With the emerging of DHIs, the existing behavioral theories can be further evaluated and improved [28]. Here we list some guidelines when using specific behavioral theories: [12] and [31] for the SCT, [8] for the HBM, [3] for the TBP, and [52] for the HAPA.

## Digital Health Intervention Taxonomy

While behavioral theories can predict and explain human behavior, there is a gap between theories and operational interventions. Will self-monitoring increases self-efficacy for promoting physical activity? Will information about health consequences affects perceived advantages/disadvantages? Due to the high complexity of human behavior and health, one DHI may involve several techniques. The lack of a consistent taxonomy of DHIs will lead to poor replicability and low comparability of the results from related studies. Although there exist taxonomies to bridge the theory-intervention gap, the use of different taxonomies still hinders the understanding and contribution among communities. Therefore, we present the DHI taxonomy, a unified taxonomy to take advantage of two widely used taxonomies (the BCT taxonomy and PSD principles) in light of the BIT model.

Behavior change techniques (BCTs) are defined as observable, replicable, and irreducible components of an intervention designed to change behavior [1,34], e.g., self-monitoring or goal setting. Abraham and Michie developed a taxonomy of behavior change techniques, which identified 22 BCTs and 4 BCT packages [1] and was later extended to a taxonomy containing 93 BCTs into 16 groups, called Behavior Change Technique Taxonomy (v1) [34]. The BCT taxonomy has been used for informing intervention development and report [37,38] and identifying the effectiveness of BCTs [17,19,33,43]. It also provides a means to evaluate health and fitness apps [13,14,16,35] and wearables [30]. From the official website of the BCT taxonomy [7], we found a collection of 405 intervention studies with BCTs coding. We show the word cloud of BCTs based on this collection in Figure 3. The top-five used or tested BCTs are goal setting (behavior), instruction on how to perform a behavior, problem-solving, information about health consequences, and action planning.

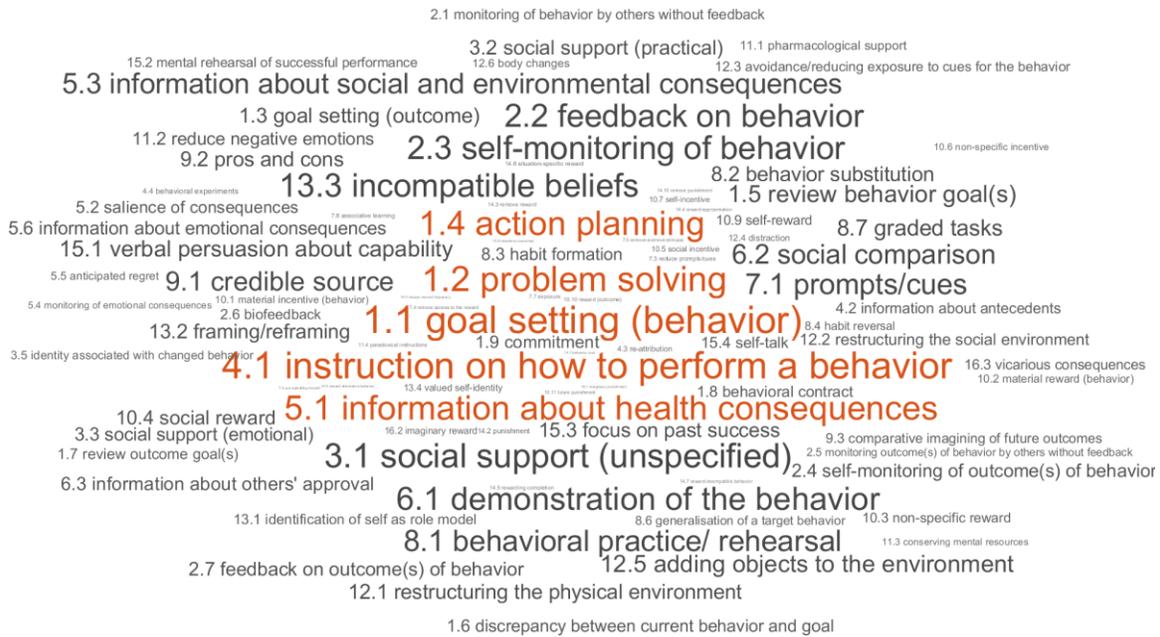

Figure 3: The word cloud of BCTs used in the study collection of 405 intervention studies from the official website of the BCT taxonomy [7].

In related work, we have introduced the behavioral intervention technology (BIT) model [36]. In terms of the intervention strategies in the BIT model, only some examples (i.e., education, goal setting, monitoring, feedback, and motivation enhancement) were provided. We think the BCT taxonomy can serve as a strategy pool for the BIT model.

Aiming to create a conceptual framework that can be directly applied to persuasive system development, the persuasive system design (PSD) model describes 28 principles in four categories (supporting primary task, computer-human dialogue, system credibility, and social) as an extension of Fogg's work on persuasive technology [18]. Table 1 describes the details of PSD principles. We found there are 16 principles (highlighted in red in Table 1) that have the same or similar definitions with counterparts from the BCT taxonomy. For example, self-monitoring appears both in PSD principles and the BCT taxonomy. Tunneling (1.2) in PSD principles has the same meaning with the BCT "4.1 structure on how to perform the behavior". Please refer to Multimedia Appendix 1 for more details. We could not find their counterparts from the BCT taxonomy of five PSD principles (highlighted in blue in Table 1), which can serve as a supplement of the BCT taxonomy.

Table 1: PSD principles. The red principles have counterparts with the same or similar definitions in the BCT taxonomy. The blue principles have no counterparts in the BCT taxonomy but can also be regarded as intervention strategies. The green principles are characteristics of intention media.

| PSD principle | Definition |
| --- | --- |
| **Primary Task Support** | |
| Reduction (1.1) | System should reduce steps users take when performing target behavior. |
| Tunneling (1.2) | System should guide users in attitude/ behavior change process by providing means for action. |
| Tailoring (1.3) | System should provide tailored info for user groups. |
| Personalization (1.4) | System should offer personalized content and services for individual users. |
| Self-monitoring (1.5) | System should provide means for users to track their performance or status. |
| Simulation (1.6) | System should provide means for observing link between cause & effect with regard to users' behavior. |
| Rehearsal (1.7) | System should provide means for rehearsing target behavior. |
| **Dialogue Support** | |
| Praise (2.1) | System should use praise to provide user feedback based on behaviors. |
| Rewards (2.2) | System should provide virtual rewards for users to give credit for performing target behavior. |
| Reminders (2.3) | System should remind users of their target behavior while using the system. |
| Suggestion (2.4) | System should suggest users carry out behaviors while using the system. |
| Similarity (2.5) | System should imitate its users in some specific way. |
| Liking (2.6) | System should have a look & feel that appeals to users. |
| Social role (2.7) | System should adopt a social role. |
| **System Credibility Support** | |
| Trust-worthiness (3.1) | System should provide info that is truthful, fair & unbiased. |
| Expertise (3.2) | System should provide info showing knowledge, experience & competence. |
| Surface credibility (3.3) | System should have competent and truthful look & feel. |
| Real-world feel (3.4) | System should provide info of the organization/actual people behind it content & services. |
| Authority (3.5) | System should refer to people in the role of authority. |
| Third-party endorsements (3.6) | System should provide endorsements from external sources. |
| Verifiability (3.7) | System should provide means to verify accuracy of site content via outside sources. |
| **Social Support** | |
| Social learning (4.1) | System should provide means to observe others performing their target behaviors. |
| Social comparison (4.2) | System should provide means for comparing performance with the performance of others. |
| Normative influence (4.3) | System should provide means for gathering people who have same goal & make them feel norms. |
| Social facilitation (4.4) | System should provide means for discerning others who are performing the behavior. |
| Cooperation (4.5) | System should provide means for cooperation. |
| Competition (4.6) | System should provide means for competing with others. |
| Recognition (4.7) | System should provide public recognition for users who perform their target behavior. |

Next, we present the diagram of our DHI taxonomy (see Figure 4). We just have shown its strategy part, which includes 93 (from the BCT taxonomy) plus 5 (real-world feel, verifiability, cooperation, competition, and recognition from PSD principles) strategies. The other part of our DHI taxonomy corresponds to the characteristics. The BIT model described four characteristics (*medium, complexity, aesthetics, and personalization*). Inspired by the characteristics related PSD principles (highlighted in green in Table 1), we include *social role* and *trustiness*, in addition to the mentioned four from the BIT model, into the characteristics part of the DHI taxonomy.

We divided the PSD principles into two groups. The ones fitting the definition of the BCT go to the strategies group, while others fall into characteristics group. Personalization is one of the characteristics in the BIT model. We find that tailoring has very close meaning to personalization according to their definitions in the PSD principles [41]. We argue that similarity is also in line with the definition of personality. Therefore, we regard both tailoring and similarity the same as personality. Likewise, trust-worthiness and surface credibility are merged to one characteristic as trustiness. By dividing the PSD principles and merging the overlapping principles, we hope our new taxonomy can reduce the confusion and difficulty of coding DHIs [21,58].

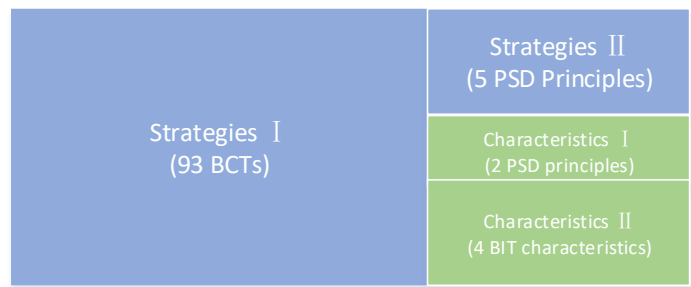

Figure 4: The diagram of our DHI taxonomy. The blue part is the strategy part, while the green part is the characteristics part.

## The Holistic Framework

The proposed holistic framework (see Figure 5) is called TUDER (Targeting, Understanding, Designing, Evaluating and Refining), which consists of four steps, two toolboxes (behavioral theories and the DHIs taxonomy), and a workflow. In each step, it is allowed to go back and update corresponding information.

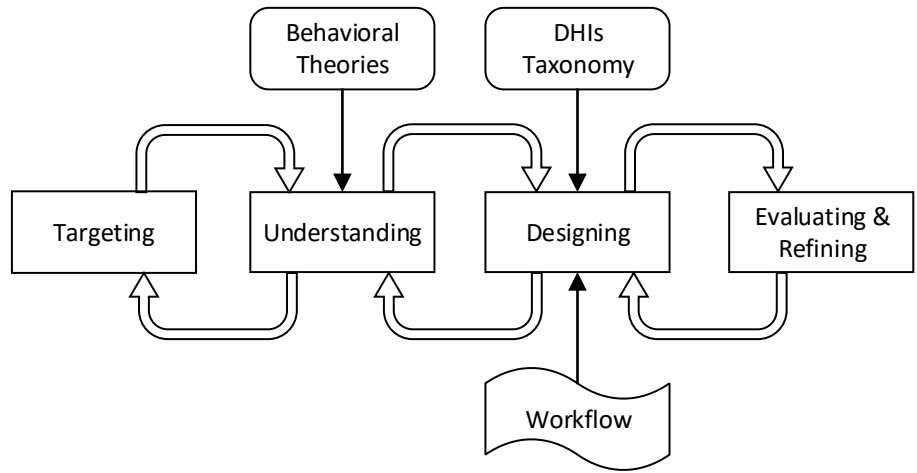

Figure 5: The diagram of TUDER (Targeting, Understanding, Designing, Evaluating and Refining).

*Targeting the user group, the health problem, and the behavior.* The target group, health problem, and behavior define the intervention aim(s). For example, an intervention to promote the use of standing desks (the behavior) to reduce the prolonged sedentary behavior (the behavior) of office workers (the user group) to prevent chronic diseases, e.g., type 2 diabetes (the health problem) [9]. The intervention designers should explain the relationship between the health problem and the behavior. Scientific evidence provides the rationale. E.g., the evidence that sedentary behavior and moderate-to-vigorous physical activity are independently associated with clustered cardiometabolic health supports the development of interventions to reduce office workers' sedentary behavior [29]. Another example is about myopia among children. A study showed that the time of outdoor activities was the most significant factor of myopia in 6- and 7-year-old Chinese children[47]. Therefore, a reasonable intervention to reduce myopia (the health problem) among children (the user group) would be increasing their outdoor activity time (the behavior). Besides the scientific support, another rule is about the measurability to enable quantitative analysis. The target health problem is not necessarily measurable in an intervention study, while the target behavior must be [59].

*Understanding the mechanism underlining human behavior.* Behavioral theories (e.g., see Figure 1 and Figure 2) provide DHI developers a toolbox to understand human behavior. Given the target user group, health problem, and behavior, developers ought to take one behavioral theory or a set of constructs from behavioral theories as the base of intervention design in the following step. We suggest that theory-based interventions should relate their strategies to specific constructs from behavioral theories. For example, an intervention design based on HAPA intended to support action planning (the construct) to reduce users' sedentary behavior [57]. Therefore, in addition to measuring the sedentary behavior, the constructs in HAPA should also be assessed. When analyzing the invention effect on action planning, the assessment of action planning is enough. However, in the case of analyzing the intervention effect on sedentary behavior, other constructs besides action planning have also to be considered. The participants should be grouped based on the level of their intention in data analysis. Alternatively, the user group in the previous step can be adjusted to only focus on one user group with a specific level of intention. During this step, DHI developers may backtrack to the previous step to adjust the target user group and measurements.

*Designing the intervention strategies, characteristics, and workflow*. We have included 98 intervention strategies and six characteristics in our DHI taxonomy. DHI developers can select a set of strategies based on their idea and describe the characteristics of their strategies according to the DHI taxonomy. As the context of an intervention may vary over time, the workflow that allows an intervention to be delivered according to time, task, or event would be demanding [36]. The workflow design has been comprehensively illustrated in the BIT model [36] and the Just-in-Time Adaptive Intervention (JITAI) framework [40]. From the perspective of implementation difficulty, *time-based workflow* (e.g., an hourly reminder in sedentary behavior intervention [58]) is the easiest. *Task-based* (e.g., a set of interventions delivered to a user sequentially) or *event-based* (e.g., adaptive food recommendation according to a user's previous meal) workflow requires user data input. Because of the difficulty of inquiring users' context data, the research on opportune moments for DHIs is still in the early stage [51,56].

*Evaluating and refining the intervention design*. Intervention evaluation could include usability evolution (regarding human-computer interaction) an effectiveness evaluation (regarding behavior change) in correspond to the uptake and impact of the intervention respectively [20]. Think-aloud [26] and cognitive walkthrough can be used in the early stage of ideation creation and prototype to identify the usability issues. Then a pilot study with a small number of participants would be deployed to test the feasibility of the whole study procedure. Because many interventions need field study, the pilot can also help find some unknown issues in real-world use. Finally, heuristic evaluations based on randomized controlled trials (RCTs) [39] or sequential multiple assignment randomized trials (SMARTs) [10] have to be conducted to generate powerful results. In our framework, an iterative evaluation and refinement process is adopted. Because evaluation and refinement are always intertwined with each other, we place them in one step in our framework.

## Discussion

We have described the TUDER, a holistic framework to guide digital health intervention (DHI) development. We also provide a checklist for DHI developers, as shown in Figure 6. By completing the checklist and reporting all the details of a DHI study, the data coding work in systematic reviews could be reduced much.

| Targeting | Understanding | Designing | Evaluating and Refining |
|---|---|---|---|
| Target user group:_____<br>Target disease:_______<br>Target behavior:_______ | Behavioral theories:____<br>Constructs:__________<br>Other factors:________ | Strategies:__________<br>Characteristics:______<br>Workflow:__________ | Study design:_________<br>Evaluation results:_____ |

Figure 6: The checklist for using TUDER.

We build TUDER based on several existing related work (e.g., [1,20,36,39,42]). The key contribution of this work is to embed behavioral theories, behavior change technology (BCT) taxonomy, and persuasive system design (PSD) principles into a holistic framework. We believe this framework will be beneficial to each of them. This holistic framework and the DHI taxonomy will also enable more research questions. We provide some examples as follows:

(1) What combinations of DHI strategies, characteristics, and workflow work better than others? In [62], a meta-analysis shows several combinations of PSD principles were more effective, e.g., tunneling and tailoring, reminders and similarity, social learning and comparison. With consideration of the characteristics and workflow when coding the interventions, the results of intervention effectiveness analysis may change.

(2) Is the DHI taxonomy able to explain more variance in DHI adherence? Kelders et al. [27] systematically reviewed the impact of the PSD principles on adherence to web-based interventions. Their model explained 55% of the variance in users' adherence. The DHI taxonomy brings more perspectives to analyze the effects of the components in interventions.

As the TUDER framework is expected to be comprehensive, some parts are simplistic. For example, only several behavioral theories are discussed. The DHI taxonomy is built upon two existing taxonomies. The DHI developers who are not familiar with the BCT taxonomy and PSD principle will find it challenging to use the DHI taxonomy.

## Conclusion

This work presented the TUDER framework, containing four steps (targeting, understanding, designing, evaluating and refining), two toolboxes (behavioral theories and digital health intervention taxonomy), and a workflow. The framework aims to integrate the advantages of behavioral theories, behavior change technique taxonomy, and persuasive technology design principles. Thus, it can help the digital health intervention researchers to design, evaluate, and report their studies in a formative and comprehensive way. By using this framework, future systematic reviews could have broader insights into digital health intervention studies. To better bridge the research from different communities, we will continue to test and improve this framework.

## Acknowledgments

This work was partially supported by SMARTACT. Y. Wang acknowledges the financial support from the China Scholarship Council (CSC).

## Appendix

Multimedia Appendix 1: The overlapping between PSD principles and the BCT taxonomy.